\documentclass[10pt, twocolumn]{article}
\usepackage[utf8]{inputenc}
\usepackage{geometry}
\usepackage{mwe}
\usepackage{afterpage}
\usepackage{amsmath}
\usepackage{amsfonts}
\usepackage{hyperref}
\usepackage{subcaption}

\usepackage{tikz}
\usepackage{xfrac}
\usepackage{siunitx}
\usepackage{float}
\usepackage{pgfplots}
\usepackage{tikz-3dplot}
\usetikzlibrary{calc}
\usetikzlibrary{math}
\pgfplotsset{width=7cm,compat=1.8}

\usepackage{color}
\usepackage{soul}

\usepackage{xcolor}

\usepackage{caption}
\usepackage[none]{hyphenat}

\makeatletter
\renewcommand{\fnum@figure}{Fig. \thefigure}
\makeatother

\usepackage{sectsty} 
\usepackage{titlesec}
\sectionfont{\fontsize{10}{12}\selectfont}
\subsectionfont{\fontsize{10}{12}\selectfont}
\titlespacing\section{0pt}{10pt plus 0pt minus 0pt}{10pt plus 0pt minus 0pt}
\titlespacing\subsection{0pt}{0pt plus 0pt minus 0pt}{0pt plus 0pt minus 0pt}
\usepackage{indentfirst} 
\titlelabel{\thetitle.\quad}

\usepackage{changepage}

\usepackage{etoolbox} 
\makeatletter
\patchcmd{\@maketitle}{\LARGE}{\fontsize{18}{24}\selectfont}{}{}
\makeatother

\usepackage[affil-it]{authblk} 

\renewcommand\Affilfont{\bfseries\fontsize{12}{14.4}\selectfont}
\makeatletter
\renewcommand\AB@authnote[1]{\textsuperscript{\normalfont\bfseries#1}}
\makeatother
\makeatletter
\renewcommand\AB@affilsepx{, \protect\Affilfont}
\makeatother

\usepackage{fancyhdr}
\usepackage[style=ieee, 
citestyle=numeric-comp,
sorting=none]{biblatex}
\usepackage{enumitem}
\setlist[itemize]{nosep}
\setlist[enumerate]{nosep}

\usepackage[symbol]{footmisc}
\renewcommand{\thefootnote}{\fnsymbol{footnote}}

\newcommand\blfootnote[1]{%
  \begingroup
  \renewcommand\thefootnote{}\footnote{#1}%
  \addtocounter{footnote}{-1}%
  \endgroup
}

\title{\textbf{Vehicle Models and Optimal Control on a Nonplanar Surface}}
\author[1]{Thomas Fork}
\author[2]{H. Eric Tseng}
\author[1]{Francesco Borrelli}
\affil[1]{UC Berkeley}
\affil[2]{Ford Research \& Advanced Engineering}
\date{\vspace{-8.5ex}}

\begin{document}

\twocolumn[
  \begin{@twocolumnfalse}
    \maketitle
    
    \begin{center}
    Email: \href{mailto:fork@berkeley.edu}{\color{blue} \underline{\smash{fork@berkeley.edu}}}
    \end{center}
    
    \begin{adjustwidth}{15mm}{15mm}
    We present a 10 DoF dynamic vehicle model for model-based control on nonplanar road surfaces. A parametric surface is used to describe the road surface, allowing the surface parameterization to describe the pose of the vehicle. We use the proposed approach to compute minimum-time vehicle trajectories on nonplanar surfaces and compare planar and nonplanar models. 
    \end{adjustwidth}
    
    \begin{center}
    Topics: Vehicle Dynamics Theory, Modeling
    \end{center}
    
  \end{@twocolumnfalse}
]

\thispagestyle{fancy} 
\renewcommand{\headrulewidth}{0pt} 

\section{BACKGROUND}
Model-based motion planning and control is widely used for vehicles~\cite{7490340}. However, commonly used models are limited to flat surfaces, limiting applicability of related control techniques. In \cite{fork2021models} we developed an approach for modeling vehicles on arbitrary nonplanar surfaces and focused on kinematic models. In this paper we leverage our approach to develop a 10 DoF dynamic vehicle model for nonplanar model-based control and use our model to compute minimum time trajectories on nonplanar surfaces. 
\blfootnote{A video can be found at \href{https://youtu.be/4nCYGlKpd2A}{https://youtu.be/4nCYGlKpd2A}}
\blfootnote{Source code: \\\href{https://github.com/thomasfork/Nonplanar-Vehicle-Control}{https://github.com/thomasfork/Nonplanar-Vehicle-Control}}

The contributions of this paper are:
\begin{enumerate}
    \item We develop a nonplanar dynamic vehicle model which includes tire forces and dynamic weight distribution.
    \item We use our dynamic vehicle model for optimal control on a nonplanar road surface.
    \item We compare our model to other nonplanar and planar vehicle models. 
\end{enumerate}

This paper is structured as follows: In sections \ref{sec:nonplanar_modeling} and \ref{sec:surface_parameterization} we introduce our vehicle model. We then use this model for optimal control in section \ref{sec:optimal_control}. We present results and conclusions in sections \ref{sec:results} and \ref{sec:conclusion}.

\section{NONPLANAR VEHICLE MODELING} \label{sec:nonplanar_modeling}
We use the frames of reference, variables and modeling approach of \cite{fork2021models}\footnote[1]{The reader may find it useful to read section II thereof.}: we model a vehicle as a rigid body in tangent contact with a parametric surface and use the surface parameterization to describe the pose of the vehicle. This requires several key assumptions:
\begin{itemize}
    \item The vehicle is treated as a single rigid body
    \item This body remains tangent to the road as it moves
    \item Contact with the road is never lost
    \item Out of plane road curvature is small relative to the length of the vehicle
\end{itemize}
We develop our model in five steps: 
\begin{enumerate}
    \item We describe the road with a parametric surface
    \item We relate vehicle velocity to motion on this surface
    \item Rigid body equations of motion are introduced
    \item Tire forces and other forces are introduced
    \item Weight distribution effects are added
\end{enumerate}
These steps are elaborated in the following subsections.

\subsection{Parametric Road Surface}
\begin{figure}
    \centering
    \begin{tikzpicture}[inner sep=0pt]
        \begin{axis}[axis line style={draw=none},
              view={-65}{35},
              tick style={draw=none},
              xticklabels={,,},
              yticklabels={,,},
              zticklabels={,,},
              clip=false
                ]
        \addplot3[
          surf,
          colormap = {graymap}{color = (lightgray) color = (lightgray)},
          shader=faceted,
          samples = 21,
          domain=-1:1,
          domain y=-1:1
        ] 
        {exp(-x^2-y^2)};
    
        \addplot3[
          black,
          mark = none,
          very thick,
          domain=-1:-0.2,
          samples y = 0
          ] 
        ({x},
         {-1},
         {exp(-x^2-1});
    
        \addplot3[
          black,
          mark = none,
          very thick,
          domain=-1:0.5,
          samples y = 0
        ] 
        ({-1},
         {x},
         {exp(-x^2-1)});
     
        \addplot3[
          black,
          mark = none,
          domain=-1:-0.2,
          samples y = 0
        ]
        ({x},
         {0.5},
         {exp(-x^2-0.25});
     
        \addplot3[
          black,
          mark = none,
          domain=-1:0.5,
          samples y = 0
        ] 
        ({-0.2},
         {x},
         {exp(-x^2-0.04)});
         
         \addplot3[
          black,
          mark = none,
          domain=0:.06,
          samples y = 0
        ] 
        ({-0.2 -0.2330* x},
         {0.5 + 0.5826 * x},
         {exp(-0.25-0.04) + 0.7786*x});
        
        \tikzmath{\h = .06;
                  \xn = -0.233;
                  \yn = 0.5826;
                  \zn = 0.7886;
                  \xc = -0.2 \xn * \h; 
                  \yc = 0.5 + \yn * \h; 
                  \zc = 0.7483 + \zn * \h;
                  \xs = 0.958;
                  \ys = 0;
                  \zs = 0.2867;
                  \xy = 0;
                  \yy = 0.801;
                  \zy = -0.5991;
                  \lf = .1;
                  \wh = .2;
                  \xlll = \xc + \xs * \lf + \xy * \wh + \xn * \h;
                  \ylll = \yc + \ys*\lf + \yy*\wh + \yn*\h;
                  \zlll = \zc + \zs*\lf + \zy*\wh + \zn*\h;
                  \xllo = \xc + \xs*\lf + \xy*\wh - \xn*\h;
                  \yllo = \yc + \ys*\lf + \yy*\wh - \yn*\h;
                  \zllo = \zc + \zs*\lf + \zy*\wh - \zn*\h;
                  \xlol = \xc + \xs*\lf - \xy*\wh + \xn*\h;
                  \ylol = \yc + \ys*\lf - \yy*\wh + \yn*\h;
                  \zlol = \zc + \zs*\lf - \zy*\wh + \zn*\h;
                  \xoll = \xc - \xs*\lf + \xy*\wh + \xn*\h;
                  \yoll = \yc - \ys*\lf + \yy*\wh + \yn*\h;
                  \zoll = \zc - \zs*\lf + \zy*\wh + \zn*\h;
                  \xloo = \xc + \xs*\lf - \xy*\wh - \xn*\h;
                  \yloo = \yc + \ys*\lf - \yy*\wh - \yn*\h;
                  \zloo = \zc + \zs*\lf - \zy*\wh - \zn*\h;
                  \xolo = \xc - \xs*\lf + \xy*\wh - \xn*\h;
                  \yolo = \yc - \ys*\lf + \yy*\wh - \yn*\h;
                  \zolo = \zc - \zs*\lf + \zy*\wh - \zn*\h;
                  \xool = \xc - \xs*\lf - \xy*\wh + \xn*\h;
                  \yool = \yc - \ys*\lf - \yy*\wh + \yn*\h;
                  \zool = \zc - \zs*\lf - \zy*\wh + \zn*\h;
                  \xooo = \xc - \xs*\lf - \xy*\wh - \xn*\h;
                  \yooo = \yc - \ys*\lf - \yy*\wh - \yn*\h;
                  \zooo = \zc - \zs*\lf - \zy*\wh - \zn*\h;}
        
        \tikzmath{\wl = -0.8;
                  \vl = 0.13;
                  \al = 0.95;
                  \xl = \xc -\xn*\h;
                  \yl = \yc -\yn*\h;
                  \zl = \zc -\zn*\h;
                  \xll = \xc - \wl*\xy;
                  \yll = \yc - \wl*\yy;
                  \zll = \zc - \wl*\zy;
                  \xlh = \xl - \wl*\xy;
                  \ylh = \yl - \wl*\yy;
                  \zlh = \zl - \wl*\zy;
                  \xllp = \xc - \wl*\xy*\al;
                  \yllp = \yc - \wl*\yy*\al;
                  \zllp = \zc - \wl*\zy*\al;
                  \xlhp = \xl - \wl*\xy*\al;
                  \ylhp = \yl - \wl*\yy*\al;
                  \zlhp = \zl - \wl*\zy*\al;
                  \xlln = \xllp + \vl*\xn;
                  \ylln = \yllp + \vl*\yn;
                  \zlln = \zllp + \vl*\zn;
                  \xlhn = \xlhp - \vl*\xn;
                  \ylhn = \ylhp - \vl*\yn;
                  \zlhn = \zlhp - \vl*\zn;}
        
        \node[circle, fill, minimum size = 4pt] (xp) at (axis cs: \xc,\yc,\zc){};
        \coordinate (xlll) at (axis cs: \xlll,\ylll,\zlll){};
        \coordinate (xoll) at (axis cs: \xoll,\yoll,\zoll){};
        \coordinate (xlol) at (axis cs: \xlol,\ylol,\zlol){};
        \coordinate (xllo) at (axis cs: \xllo,\yllo,\zllo){};
        \coordinate (xool) at (axis cs: \xool,\yool,\zool){};
        \coordinate (xolo) at (axis cs: \xolo,\yolo,\zolo){};
        \coordinate (xloo) at (axis cs: \xloo,\yloo,\zloo){};
        \coordinate (xooo) at (axis cs: \xooo,\yooo,\zooo){};
        \begin{scope}[line join=round, fill opacity = 0.2]
            \draw[fill=green] (xlll) -- (xlol) -- (xloo) -- (xllo) -- cycle ; 
            \draw[fill=green] (xllo) -- (xolo) -- (xooo) -- (xloo) -- cycle ;  
            \draw[fill=green] (xlll) -- (xoll) -- (xolo) -- (xllo) -- cycle ; 
            \draw[fill=green] (xlol) -- (xool) -- (xooo) -- (xloo) -- cycle ; 
            \draw[fill=green] (xoll) -- (xool) -- (xooo) -- (xolo) -- cycle ;
            \draw[fill=green] (xlll) -- (xoll) -- (xool) -- (xlol) -- cycle ;  
        \end{scope}

        \coordinate (xl) at (axis cs: \xl,\yl,\zl);
        \coordinate (xlh) at (axis cs: \xlh,\ylh,\zlh);
        \coordinate (xll) at (axis cs: \xll,\yll,\zll);
        \coordinate (xlhp) at (axis cs: \xlhp,\ylhp,\zlhp);
        \coordinate (xllp) at (axis cs: \xllp,\yllp,\zllp);
        \coordinate (xlhn) at (axis cs: \xlhn,\ylhn,\zlhn);
        \coordinate (xlln) at (axis cs: \xlln,\ylln,\zlln);
        
        \begin{scope}[line join=round, thin]
            \draw (xp) -- (xll);
            \draw (xl) -- (xlh);
            
            \draw[->] (xlln) -- (xllp);
            \draw[->] (xlhn) -- (xlhp);
        \end{scope}
        \node[left= 0.1cm, below = 0.3cm] (n)  at (xlln){\scalebox{0.7}{$n$}};

        \coordinate (A)  at (axis cs: -1, -1, 0.1353){};
        \coordinate[below = 0.4cm, left = 0.2cm] (Ay) at (A);
        \coordinate[below = 0.2cm, left = 0.1cm] (Ayh) at (A);
        \coordinate (B)  at (axis cs: -1, 0.5, 0.2865);
        \coordinate[below = 0.4cm, left = 0.2cm] (By) at (B);
        \coordinate[below = 0.2cm, left = 0.1cm] (Byh) at (B);
        \draw[very thin] (A) -- (Ay);
        \draw[very thin] (B) -- (By);
        \draw[bend right=12, ->] (Ayh) to (Byh);
        \node[fill=white,rounded corners=2pt, inner sep = 2pt] (y) at (axis cs: -1, -0.15, 0.2){\scalebox{0.7}{$y$}};
        \coordinate[below = 0.3cm, right = 0.4cm] (As) at (A);
        \coordinate[below = 0.15cm, right = 0.2cm] (Ash) at (A);
        \coordinate (C)  at (axis cs: -0.2,-1,0.3534);
        \coordinate[below = 0.4cm, right = 0.4cm] (Cs)  at (C);
        \coordinate[below = 0.2cm, right = 0.2cm] (Csh)  at (C);
        \draw[very thin] (A) -- (As);
        \draw[very thin] (C) -- (Cs);
        \draw[bend left=8, ->] (Ash) to (Csh);
        \node[fill=white,rounded corners=2pt, inner sep = 2pt] (s) at (axis cs: -0.4, -1,0.07){\scalebox{0.7}{$s$}};

        \end{axis}
    \end{tikzpicture}
    
    \caption{Parametric road surface and variables}
    \label{fig:s_y_n}
\end{figure}
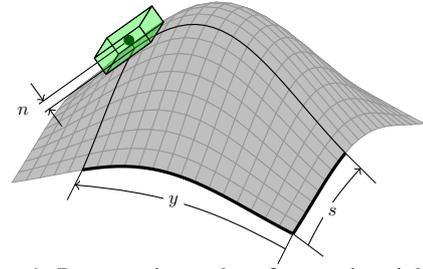

We describe the road surface with a known parametric surface $\boldsymbol{x}^p(s,y)$ where $\boldsymbol{x}^p$ is a vector in $\mathbb{R}^3$ in an inertial frame of reference and $s$ and $y$ are parameterization variables (often, but not necessarily path length and lane offset respectively). We assume that this road surface is smooth, time invariant and regular, meaning that the tangent vectors of the surface are linearly independent. We extend \cite{fork2021models} in adding a third parametric variable: $n$, to offset the vehicle center of mass (COM) normal to the parametric surface\footnote[7]{Without this, in \cite{fork2021models}, it was necessary to use a parametric surface that always contained the center of mass of the vehicle: an offset of the true road surface.}. This is illustrated in Figure \ref{fig:s_y_n}.

We denote the tangent vectors with the partial derivatives $\frac{\partial}{\partial s} \boldsymbol{x}^p = \boldsymbol{x}^p_s$ and $\frac{\partial}{\partial y} \boldsymbol{x}^p = \boldsymbol{x}^p_y$. We define the outward normal vector $\boldsymbol{e}^p_n$ as
\begin{equation}
    \boldsymbol{e}^p_n = \frac{\boldsymbol{x}^p_s \times \boldsymbol{x}^p_y }{||\boldsymbol{x}^p_s \times \boldsymbol{x}^p_y ||}. 
\end{equation}

\begin{figure}
    \centering
    \begin{tikzpicture}[inner sep=0pt]

    \node[anchor = west] (g1) at (.55,0){\scalebox{0.7}{$\boldsymbol{e}_1^g$}};
    \node[anchor = west] (g2) at (.4,.4){\scalebox{0.7}{$\boldsymbol{e}_2^g$}};
    \node[anchor = west] (g3) at (-0.1,.65){\scalebox{0.7}{$\boldsymbol{e}_3^g$}};
    \draw[thick,->](0,0)--(0,.5);
    \draw[thick,->](0,0)--(0.35,.35);
    \draw[thick,->](0,0)--(.5,0);
    
    \begin{axis}[axis line style={draw=none},
          view={35}{75},
          tick style={draw=none},
          xticklabels={,,},
          yticklabels={,,},
          zticklabels={,,},
          clip=false
            ]

    \addplot3[
      surf,
      colormap = {graymap}{color = (lightgray) color = (lightgray)},
      shader=faceted,
      samples = 17,
      domain=-0.8:0,
      domain y= 0 : pi/2,
      opacity = 1
    ]
    ({sqrt(1-x^2) * cos(deg(y))},
     {sqrt( 1-x^2 ) * sin(deg(y))},
     {-x});

    \addplot3[
      black,
      mark = none,
      very thick,
      domain=0:pi/4,
      samples y = 0
    ] 
    ({sqrt(1) * cos(deg(x))},
     {sqrt(1) * sin(deg(x))},
     {0});
    
    \addplot3[
      black,
      mark = none,
      very thick,
      domain=0:-.6,
      samples y = 0
    ] 
    ({sqrt(1-x^2) * cos(deg(0))},
     {sqrt( 1-x^2 ) * sin(deg(0))},
     {-x});
     
    \addplot3[
      black,
      mark = none,
      domain=0:pi/4,
      samples y = 0
    ] 
    ({sqrt(1-.60^2) * cos(deg(x))},
     {sqrt(1-.60^2 ) * sin(deg(x))},
     {0.60});
     
    \addplot3[
      black,
      mark = none,
      domain=0:-.6,
      samples y = 0
    ] 
    ({sqrt(1-x^2) * cos(deg(pi/4))},
     {sqrt( 1-x^2 ) * sin(deg(pi/4))},
     {-x});
    
    \node[circle, fill, minimum size = 2pt] (b0) at (axis cs: 0.566, 0.566, 0.6){};
    
    \node[anchor = west] (b1) at (axis cs: 0.3, 0.7, 0.6){};
    \node[anchor = west,left = 0.18cm,below = 0.1cm,fill=white,rounded corners=2pt] (b1l) at (b1){\scalebox{0.7}{$\boldsymbol{e}_1^b$}};
    
    \node[anchor = south,fill=white,rounded corners=2pt] (b2) at (axis cs: .5, .3, 0.6){\scalebox{0.7}{$\boldsymbol{e}_2^b$}};
    
    \node[anchor = center] (xpu) at (axis cs: 0.066, 1.066, 0.6){};
    \node[anchor = west, right = 0.15cm](xps) at (xpu){\scalebox{0.7}{$\boldsymbol{x}^p_s$}};
    
    \node[anchor = west,fill=white,rounded corners=2pt] (xpv) at (axis cs: 0.266, 0.266, 0.85) {\scalebox{0.7}{$\boldsymbol{x}^p_y$}};
    
    \node(en) at (axis cs: .65, .6, 1.5){};
    \node(enl)[anchor = east, right = 0.1cm,fill=white,rounded corners=2pt] at (en){\scalebox{0.7}{$\boldsymbol{e}_n^p,\boldsymbol{e}_3^b$}};
    
    \node[circle, fill, minimum size = 4pt] (com) at ($(en)!0.5!(b0)$) {};
    \node[right = 2cm, rotate around = {-30:(b0)}] (coml1) at (b0) {};
    \node[right = 2cm, rotate around = {-30:(com)}] (coml2) at (com) {};
    
    \draw[thick,blue,->](b0)--(b1);
    \draw[thick,blue,->](b0)--(b2);
    
    \draw[thick,black,->](b0)--(xpu);
    \draw[thick,black,->](b0)--(xpv);
    \draw[postaction={draw,black,dash pattern= on 3pt off 3pt,dash phase=3pt,thick,->}][blue,dash pattern= on 3pt off 3pt,thick,->](b0)--(en);
    
    \node[] (thu1) at (xpu){};
    \node[] (thu2) at (axis cs: 0,0.85,0.7){};
    \draw[very thin,] (b0) -- (thu1);
    \draw[very thin,] (b0) -- (thu2);
    \draw[bend right, ->] (thu1) to (thu2) ;
    \node (thu) at (axis cs: 0, 1, .9) {\scalebox{0.7}{$\theta^s$}};
    \node[coordinate, pin = {[pin distance=0.5cm,pin edge = {thick,dashed,black}]left:{\scalebox{0.7}{Road Surface }}}] at (axis cs: .588, 0.119, 0.8){};
    \node[coordinate, pin = {[pin distance=1.4cm, pin edge = {thick,dashed,black}]0:{\scalebox{0.7}{ $\boldsymbol{x}^p$}}}] at (b0){};
    \node[coordinate, pin = {[pin distance=1.4cm, pin edge = {thick,dashed,black}]0:{\scalebox{0.7}{COM}}}] at (com){};
    \node (n_label) at ($(coml1)!0.5!(coml2)$) {\scalebox{0.7}{$n$}};
    
    \draw[very thin] (b0) -- (coml1);
    \draw[very thin] (com) -- (coml2);

    \node (A)  at (axis cs: 1, 0,0){};
    \node[below = 0.4cm, left = 0.4cm] (Ay) at (A){};
    \node[below = 0.2cm, left = 0.2cm] (Ayh) at (A){};
    \node (B)  at (axis cs: 0.8, 0,0.6){};
    \node[below = 0.4cm, left = 0.4cm] (By) at (B){};
    \node[below = 0.2cm, left = 0.2cm] (Byh) at (B){};
    \draw[very thin] (A) -- (Ay);
    \draw[very thin] (B) -- (By);
    \draw[bend right=12, ->] (Ayh) to (Byh);
    \node[fill=white,rounded corners=2pt, inner sep = 2pt] (y) at (axis cs: 0.9, -0.1,0.4){\scalebox{0.7}{$y$}};
    
    \node (As) at (axis cs: 1.2,-0,0){};
    \node (Ash) at (axis cs: 1.1,-0,0){};
    \node (C)  at (axis cs: 0.707,0.707,0){};
    \node (Cs)  at (axis cs: .85,.85,0){};
    \node (Csh)  at (axis cs: 0.77,0.77,0.0){};
    \draw[very thin] (A) -- (As);
    \draw[very thin] (C) -- (Cs);
    \draw[bend right, ->] (Ash) to (Csh);
    \node[fill=white,rounded corners=2pt, inner sep = 2pt] (s) at (axis cs: 1.025, 0.40,0){\scalebox{0.7}{$s$}};

    \end{axis}

    \end{tikzpicture}
    \caption{Coordinate systems and $\theta^s$}
    \label{fig:variables}
\end{figure}
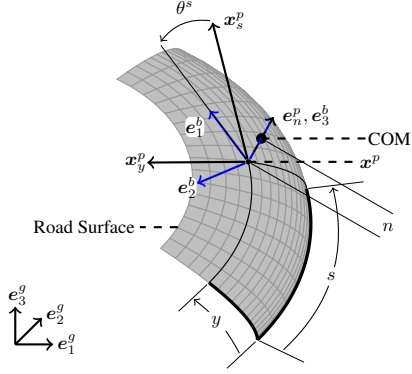

Variables $s$ and $y$ describe position on the surface; to describe orientation we introduce $\theta^s$ as shown in Figure \ref{fig:variables}, with the mathematical definition below:
\begin{subequations}\label{eq:parametric_angle}
\begin{equation}
    \cos(\theta^s) = \frac{\boldsymbol{e}_1^b  \cdot \boldsymbol{x}^p_s}{||\boldsymbol{x}^p_s||}
\end{equation}
\begin{equation}
    \sin(\theta^s) = \frac{-\boldsymbol{e}_2^b \cdot \boldsymbol{x}^p_s}{||\boldsymbol{x}^p_s||}.
\end{equation}
\end{subequations}
Imposing tangent contact means imposing the constraints
\begin{subequations} \label{eq:tangent_contact_constraints}
    \begin{equation} \label{eq:surface_constraint}
        \boldsymbol{x}^p + n\boldsymbol{e}^p_n = \boldsymbol{x}^g
    \end{equation}
    \begin{equation} \label{eq:tangent_constraint_1}
        \boldsymbol{e}_3^b \cdot \boldsymbol{x}^p_s = 0
    \end{equation}
    \begin{equation} \label{eq:tangent_constraint_2}
        \boldsymbol{e}_3^b \cdot \boldsymbol{x}^p_y = 0.
    \end{equation}
\end{subequations}
where $\boldsymbol{x}^g$ is position in the global, inertial frame of reference. 

\subsection{Parametric Motion Equations} \label{sec:surface_motion_eqns}
We first relate linear and angular velocity of the vehicle in the body frame to the time derivative of vehicle pose: $\dot{s}$, $\dot{y}$, $\dot{n}$ and $\dot{\theta}^s$, providing the first set of equations for our vehicle model. As in \cite{fork2021models}, we do so by differentiating constraints \eqref{eq:parametric_angle} and \eqref{eq:tangent_contact_constraints} with respect to time. Only \eqref{eq:surface_constraint} has changed, so we reuse the results of \cite{fork2021models} for the remaining equations:

\begin{equation} \label{eq:surface_angular_velocity_unsimplified}
    \begin{split}
        \underbrace{
        \begin{bmatrix}
        \boldsymbol{x}^p_{ss} \cdot \boldsymbol{e}_n^p & \boldsymbol{x}^p_{sy} \cdot \boldsymbol{e}_n^p \\
        \boldsymbol{x}^p_{ys} \cdot \boldsymbol{e}_n^p & \boldsymbol{x}^p_{yy} \cdot \boldsymbol{e}_n^p
        \end{bmatrix}
        }_{\mathbf{II}}
        \begin{bmatrix}
        \dot{s} \\ \dot{y}
        \end{bmatrix}
        =
        \underbrace{
        \begin{bmatrix}
        \boldsymbol{x}^p_s \cdot \boldsymbol{e}_1^b & \boldsymbol{x}^p_s \cdot \boldsymbol{e}_2^b \\
        \boldsymbol{x}^p_y \cdot \boldsymbol{e}_1^b & \boldsymbol{x}^p_y \cdot \boldsymbol{e}_2^b
        \end{bmatrix}
        }_{\mathbf{J}}
        \begin{bmatrix}
        -\omega_2^b \\ \omega_1^b
        \end{bmatrix}
    \end{split}
\end{equation}

\begin{equation} \label{eq:track_angular_velocity}
    \begin{split}
        \dot{\theta}^s = \omega_3^b + 
          \frac{\left(\boldsymbol{x}^p_{ss}\times \boldsymbol{x}^p_s\right)\cdot \boldsymbol{e}_n^p}{\boldsymbol{x}^p_s \cdot \boldsymbol{x}^p_s}\dot{s} 
        \\
        + \frac{\left(\boldsymbol{x}^p_{sy}\times \boldsymbol{x}^p_s\right)\cdot \boldsymbol{e}_n^p}{\boldsymbol{x}^p_s \cdot \boldsymbol{x}^p_s}\dot{y}.
    \end{split}
\end{equation}
$\boldsymbol{x}^p_{ss}$ is our notation for a second partial derivative. Recognizing that $\frac{d}{dt}\boldsymbol{e}^p_n = \frac{d}{dt}\boldsymbol{e}^b_3 = \omega_2^b \boldsymbol{e}^b_1 - \omega^b_1 \boldsymbol{e}^b_3$, the time derivative of \eqref{eq:surface_constraint} is:
\begin{equation}
\begin{split}
    \dot{s}\boldsymbol{x}^p_s + \dot{y}\boldsymbol{x}^p_y + \dot{n} \boldsymbol{e}^p_n +n(\omega_2^b \boldsymbol{e}^b_1 -\omega^b_1 \boldsymbol{e}^b_2) \\= v^b_1 \boldsymbol{e}^b_1 + v^b_2 \boldsymbol{e}^b_2 + v^b_3 \boldsymbol{e}^b_3
\end{split}
\end{equation}
which we simplify by taking inner products with respect to $\boldsymbol{x}^p_s$, $\boldsymbol{x}^p_y$ and $\boldsymbol{e}^p_n$:
\begin{equation} \label{eq:parametric_velocity_unsimplified}
    \begin{split}
        \underbrace{
        \begin{bmatrix}
        \boldsymbol{x}^p_s \cdot \boldsymbol{e}_1^b & \boldsymbol{x}^p_s \cdot \boldsymbol{e}_2^b\\
        \boldsymbol{x}^p_y \cdot \boldsymbol{e}_1^b & \boldsymbol{x}^p_y \cdot \boldsymbol{e}_2^b
        \end{bmatrix}
        }_{\mathbf{J}}
        \left(
        \begin{bmatrix}
        v_1^b \\
        v_2^b
        \end{bmatrix}
        +n
        \begin{bmatrix}
        -\omega^b_2 \\
        \omega^b_1
        \end{bmatrix}
        \right)
        \\=
        \underbrace{
        \begin{bmatrix}
        \boldsymbol{x}^p_s \cdot \boldsymbol{x}^p_s  & \boldsymbol{x}^p_s \cdot \boldsymbol{x}^p_y \\
        \boldsymbol{x}^p_s \cdot \boldsymbol{x}^p_y & \boldsymbol{x}^p_y \cdot \boldsymbol{x}^p_y
        \end{bmatrix}
        }_{\mathbf{I}}
        \begin{bmatrix}
        \dot{s}\\
        \dot{y}
        \end{bmatrix}
    \end{split}
\end{equation}
\begin{equation}\label{eq:normal_velocity}
    \dot{n} = v^b_3.
\end{equation}

Using \eqref{eq:surface_angular_velocity_unsimplified} we simplify \eqref{eq:parametric_velocity_unsimplified} to:
\begin{equation} \label{eq:parametric_velocity_simplified}
    \begin{bmatrix}
    \dot{s}\\
    \dot{y}
    \end{bmatrix}
    =
    \left(\mathbf{I} - n \mathbf{II}\right)^{-1}
    \mathbf{J}
    \begin{bmatrix}
        v_1^b \\
        v_2^b
    \end{bmatrix}
\end{equation}
which when $n=0$ is identical to the expression obtained in \cite{fork2021models}. Together, \eqref{eq:track_angular_velocity}, \eqref{eq:normal_velocity} and \eqref{eq:parametric_velocity_simplified} relate vehicle velocity to parametric velocity $(\dot{s}, \dot{y}, \dot{n}, \dot{\theta}^s)$ and \eqref{eq:surface_angular_velocity_unsimplified} gives the necessary angular velocity $\omega^b_1$ and $\omega^b_2$ to follow a curved surface while remaining tangent to it. 

Note that combining \eqref{eq:surface_angular_velocity_unsimplified} and \eqref{eq:parametric_velocity_simplified} we get
\begin{equation} \label{eq:angular_vel_constraint}
    \begin{bmatrix}
        -\omega_2^b \\ \omega_1^b
    \end{bmatrix}
    =\mathbf{J}^{-1} \mathbf{II}\ \left(\mathbf{I} - n \mathbf{II}\right)^{-1} \mathbf{J}
    \begin{bmatrix}
    v_1^b \\
    v_2^b
    \end{bmatrix}
\end{equation}
which represents the curvature of the road surface in the body frame of the vehicle. 

In the following section $\dot{\omega}^b_1$ and $\dot{\omega}^b_2$ will impact the weight distribution of a vehicle, necessitating expressions for both. Whereas the previous expressions are exact we make the approximation:

\begin{equation} \label{eq:angular_accel_constraint}
    \begin{bmatrix}
        -\dot{\omega}_2^b \\ \dot{\omega}_1^b
    \end{bmatrix}
    =\mathbf{J}^{-1} \mathbf{II}\ \left(\mathbf{I} - n \mathbf{II}\right)^{-1} \mathbf{J}
    \begin{bmatrix}
    \dot{v}_1^b \\
    \dot{v}_2^b
    \end{bmatrix}.
\end{equation}
This assumes that the curvature of the surface changes gradually enough that the omitted terms are negligible. We claim this is the case for road surfaces of interest, as we have already assumed the road curvature is small relative to the length of the vehicle.

Henceforth we limit ourselves to the case $\dot{n} = 0$.

\subsection{Vehicle Velocity Equations}
We derive velocity equations of motion by simplifying the Newton Euler equations for our modeling approach. We use the common assumption that the moment of inertia matrix of a vehicle is diagonal~\cite[Ch.~9]{guiggiani_book}. Written out, the equations are \cite{mechanics_landau_lifshitz}:
\begin{subequations} \label{eq:ne}
    \begin{equation} \label{eq:ne_v1}
        \dot{v}^b_1 + v^b_3 \omega^b_2 - v^b_2 \omega^b_3 = \frac{1}{m}F^b_1
    \end{equation}
    \begin{equation} \label{eq:ne_v2}
        \dot{v}^b_2 + v^b_1 \omega^b_3 - v^b_3 \omega^b_1 = \frac{1}{m}F^b_2
    \end{equation}
    \begin{equation} \label{eq:ne_v3}
        \dot{v}^b_3 + v^b_2 \omega^b_1 - v^b_1 \omega^b_2 = \frac{1}{m}F^b_3
    \end{equation}
    \begin{equation} \label{eq:ne_w1}
        I^b_1 \dot{\omega}^b_1 + \left(I^b_3 - I^b_2\right)\omega^b_2 \omega^b_3 = K^b_1
    \end{equation}
    \begin{equation} \label{eq:ne_w2}
        I^b_2 \dot{\omega}^b_2 + \left(I^b_1 - I^b_3\right)\omega^b_3 \omega^b_1 = K^b_2
    \end{equation}
    \begin{equation} \label{eq:ne_w3}
        I^b_3 \dot{\omega}^b_3 + \left(I^b_2 - I^b_1\right)\omega^b_1 \omega^b_2 = K^b_3.
    \end{equation}
\end{subequations}

Equations \eqref{eq:ne_v3}, \eqref{eq:ne_w1} and \eqref{eq:ne_w2} are constrained by our tangent contact assumption and impact weight distribution. We solve the remaining parts of \eqref{eq:ne} to derive velocity equations of motion. We consider three sources of force $\boldsymbol{F}^b$ and moment $\boldsymbol{K}^b$ in \eqref{eq:ne}: gravity $\boldsymbol{F}^{g,b}$, tire forces and drag $\boldsymbol{F}^{d,b}$, $\boldsymbol{K}^{d,b}$. We assume that gravity and drag are functions of the vehicle state, ie. pose and velocity, whereas tire forces must be found by solving for the weight distribution of the vehicle as in \cite{rucco2014development}. Once we collect the vehicle pose, velocity, tire force and input variables, we obtain a differential algebraic model (DAE), which we cast in the semi-explicit form:
\begin{subequations} \label{eq:dae_general_form}
\begin{equation}
    \dot{\mathcal{Z}} = f(\mathcal{Z}, \mathcal{U}, \mathcal{G})
\end{equation}
\begin{equation}
    0 = g_c(\mathcal{Z}, \mathcal{U}, \mathcal{G})
\end{equation}
\end{subequations}
where $\mathcal{Z}$ is a state vector, $\mathcal{U}$ an input vector and $\mathcal{G}$ a vector of algebraic states. Vehicle dynamics are captured by $f$ with weight distribution enforced by $g_c$. Next we define $\mathcal{Z}$, $\mathcal{U}$ and $\mathcal{G}$ for our model and derive $f$ and $g_c$

\subsection{Two Track Vehicle Model}
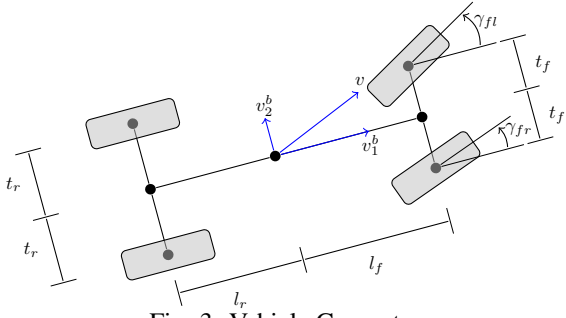
\begin{figure}
    \centering
    \begin{tikzpicture}[inner sep=0pt]
        \tikzmath{\lf = 2;
                  \lr = 1.7;
                  \tf = .7;
                  \tr = .9;
                  \th = 15;
                  \yfr = 20;
                  \yfl = 30;
                  \v1 = 1.3;
                  \v2 = .545;}
    
        \node[circle, fill, minimum size = 4pt] (c) at (0,0) {};
        \node[circle, fill, minimum size = 4pt, rotate around={\th:(c)}] (f) at (\lf,0) {};
        \node[circle, fill, minimum size = 4pt, rotate around={\th:(c)}] (r) at (-\lr,0) {};
        \node[circle, fill, minimum size = 4pt, rotate around={\th:(c)}] (fr) at (\lf,-\tf) {};
        \node[circle, fill, minimum size = 4pt, rotate around={\th:(c)}] (fl) at (\lf,\tf) {};
        \node[circle, fill, minimum size = 4pt, rotate around={\th:(c)}] (rr) at (-\lr,-\tr) {};
        \node[circle, fill, minimum size = 4pt, rotate around={\th:(c)}] (rl) at (-\lr,\tr) {};
        \draw (c) -- (r);
        \draw (c) -- (f);
        \draw (f) -- (fr);
        \draw (f) -- (fl);
        \draw (r) -- (rr);
        \draw (r) -- (rl);
        
        \node[below = 1.6cm, rotate around = (\th:(f))] (lfl) at (f) {};
        \node[below = 1.4cm, rotate around = (\th:(f))] (lfh) at (f) {};
        \node[below = 1.2cm, rotate around = (\th:(f))] (lfs) at (f) {};
        \node[below = 1.6cm, rotate around = (\th:(r))] (lrl) at (r) {};
        \node[below = 1.4cm, rotate around = (\th:(r))] (lrh) at (r) {};
        \node[below = 1.2cm, rotate around = (\th:(r))] (lrs) at (r) {};
        \node[below = 1.6cm, rotate around = (\th:(c))] (cl) at (c) {};
        \node[below = 1.4cm, rotate around = (\th:(c))] (ch) at (c) {};
        \node[below = 1.2cm, rotate around = (\th:(c))] (cs) at (c) {};
        \draw[very thin] (lfs) -- (lfl);
        \draw[very thin] (lrs) -- (lrl);
        \draw[very thin] (cs) -- (cl);
        \draw[thin] (lrh) -- (ch);
        \draw[thin] (lfh) -- (ch);
        \node[anchor = center, below = 0.2cm] (lf) at ($(lfh)!0.5!(ch)$) {\scalebox{0.7}{$l_f$}};
        \node[anchor = center, below = 0.2cm] (lr) at ($(lrh)!0.5!(ch)$) {\scalebox{0.7}{$l_r$}};
        
        \node[right = 1.6cm, rotate around = (\th:(fr))] (frl) at (fr) {};
        \node[right = 1.4cm, rotate around = (\th:(fr))] (frh) at (fr) {};
        \node[right = 1.2cm, rotate around = (\th:(fr))] (frs) at (fr) {};
        \node[right = 1.6cm, rotate around = (\th:(f))] (fcl) at (f) {};
        \node[right = 1.4cm, rotate around = (\th:(f))] (fch) at (f) {};
        \node[right = 1.2cm, rotate around = (\th:(f))] (fcs) at (f) {};
        \node[right = 1.6cm, rotate around = (\th:(fl))] (fll) at (fl) {};
        \node[right = 1.4cm, rotate around = (\th:(fl))] (flh) at (fl) {};
        \node[right = 1.2cm, rotate around = (\th:(fl))] (fls) at (fl) {};
        \draw[very thin] (frl) -- (frs);
        \draw[very thin] (fcl) -- (fcs);
        \draw[very thin] (fll) -- (fls);
        \draw[thin] (frh) -- (fch);
        \draw[thin] (flh) -- (fch);
        \node[anchor = center, right = 0.2cm] (tf1) at ($(frh)!0.5!(fch)$) {\scalebox{0.7}{$t_f$}};
        \node[anchor = center, right = 0.2cm] (tf2) at ($(flh)!0.5!(fch)$) {\scalebox{0.7}{$t_f$}};
        
        \node[left = 1.6cm, rotate around = (\th:(rr))] (rrl) at (rr) {};
        \node[left = 1.4cm, rotate around = (\th:(rr))] (rrh) at (rr) {};
        \node[left = 1.2cm, rotate around = (\th:(rr))] (rrs) at (rr) {};
        \node[left = 1.6cm, rotate around = (\th:(r))] (rcl) at (r) {};
        \node[left = 1.4cm, rotate around = (\th:(r))] (rch) at (r) {};
        \node[left = 1.2cm, rotate around = (\th:(r))] (rcs) at (r) {};
        \node[left = 1.6cm, rotate around = (\th:(rl))] (rll) at (rl) {};
        \node[left = 1.4cm, rotate around = (\th:(rl))] (rlh) at (rl) {};
        \node[left = 1.2cm, rotate around = (\th:(rl))] (rls) at (rl) {};
        \draw[very thin] (rrl) -- (rrs);
        \draw[very thin] (rcl) -- (rcs);
        \draw[very thin] (rll) -- (rls);
        \draw[thin] (rrh) -- (rch);
        \draw[thin] (rlh) -- (rch);
        \node[anchor = center, left = 0.2cm] (tr1) at ($(rrh)!0.5!(rch)$) {\scalebox{0.7}{$t_r$}};
        \node[anchor = center, left = 0.2cm] (tr2) at ($(rlh)!0.5!(rch)$) {\scalebox{0.7}{$t_r$}};
        
        \node[rectangle, fill = lightgray, draw = black, fill opacity = 0.5, anchor = center, rotate around = (\th + \yfr:(fr)),minimum width=1.2cm,minimum height=0.4cm,rounded corners=2pt] (tfr) at (fr) {};
        
        \node[rectangle, fill = lightgray, draw = black, fill opacity = 0.5, anchor = center, rotate around = (\th + \yfl:(fl)),minimum width=1.2cm,minimum height=0.4cm,rounded corners=2pt] (tfl) at (fl) {};
        
        \node[rectangle, fill = lightgray, draw = black, fill opacity = 0.5, anchor = center, rotate around = (\th:(rr)),minimum width=1.2cm,minimum height=0.4cm,rounded corners=2pt] (trr) at (rr) {};
        
        \node[rectangle, fill = lightgray, draw = black, fill opacity = 0.5, anchor = center, rotate around = (\th:(rl)),minimum width=1.2cm,minimum height=0.4cm,rounded corners=2pt] (trl) at (rl) {};
        
        \node[right = 1.2cm, rotate around = (\th+\yfr:(fr))] (yfrl) at (fr) {};
        \node[right = 1.2cm, rotate around = (\th     :(fr))] (yfrs) at (fr) {};
        \node (yfrsh) at ($(fr)!0.8!(yfrs)$) {};
        \node (yfrlh) at ($(fr)!0.8!(yfrl)$) {};
        \node[anchor = center] (yfrc) at ($(yfrl)!0.5!(yfrs)$) {\scalebox{0.7}{$\gamma_{fr}$}};
        \draw[very thin] (fr) -- (yfrl);
        \draw[very thin] (fr) -- (yfrs);
        \draw[bend right=\yfr, ->] (yfrsh) to (yfrlh);
        
        \node[right = 1.2cm, rotate around = (\th+\yfl:(fl))] (yfll) at (fl) {};
        \node[right = 1.2cm, rotate around = (\th     :(fl))] (yfls) at (fl) {};
        \node (yflsh) at ($(fl)!0.8!(yfls)$) {};
        \node (yfllh) at ($(fl)!0.8!(yfll)$) {};
        \node[anchor = center] (yflc) at ($(yfll)!0.5!(yfls)$) {\scalebox{0.7}{$\gamma_{fl}$}};
        \draw[very thin] (fl) -- (yfll);
        \draw[very thin] (fl) -- (yfls);
        \draw[bend right=\yfl, ->] (yflsh) to (yfllh);
        
        \node[rotate around = (\th:(c))] (v1) at (\v1,0) {};
        \node[rotate around = (\th:(c))] (v2) at (0,\v2) {};
        \node[rotate around = (\th:(c))] (v) at (\v1,\v2) {};
        \node (v1h) at ($(v1)!0.2!(c)$) {};
        \node (vh)  at ($(v)!0.2!(c)$)  {};
        \draw[blue, ->] (c) -- (v1);
        \draw[blue, ->] (c) -- (v2);
        \draw[blue, ->] (c) -- (v);
        \node[anchor = north, below = 0.05cm] (v1l) at (v1) {\scalebox{0.7}{$v_1^b$}};
        \node[anchor = south] (v2l) at (v2) {\scalebox{0.7}{$v_2^b$}};
        \node[anchor = west, above = 0.07cm] (vl) at (v) {\scalebox{0.7}{$v$}};
        
    \end{tikzpicture}
    \caption{Vehicle Geometry}
    \label{fig:vehicle_geometry}
\end{figure}
First we define basic vehicle geometry, as shown in Figure \ref{fig:vehicle_geometry}. We use $l_f$ and $l_r$ for the distance to the front and rear axle from the center of mass respectively. We use $t_f$ for half the width of the front axle and $t_r$ for half the width of the rear axle. We use Ackermann steering for the front steering angles $\gamma_{fr}$ and $\gamma_{fl}$ with a single front steering angle input $\gamma$. We also treat the slip ratio of each tire as an input. With this in mind, the state and input variables in \eqref{eq:dae_general_form} are:
\begin{subequations}
\begin{equation}
    \mathcal{Z} = \left[ s, y, \theta^s, v^b_1, v^b_2, \omega^b_3 \right]
\end{equation}
\begin{equation}
    \mathcal{U} = \left[\sigma_{fr}, \sigma_{fl}, \sigma_{rr}, \sigma_{rl}, \gamma\right]
\end{equation}
\end{subequations}
where $\sigma_{ij}$ is the slip ratio of a given tire with the first index corresponding to front (f) or rear (r) and the second to right (r) or left (l) side of the vehicle. A simple choice for $\mathcal{G}$ is $\mathcal{G} = \left[N_{fr}, N_{rl}, N_{rr}, N_{rl} \right]$, where $N_{ij}$ is the normal force on a given tire. We present a simpler expression for $\mathcal{G}$ in the following section. 

We model each tire with a combined slip Pacejka tire model that is linear in normal force and based on \cite[pgs. 179-182]{pacejka_tire_book}. Using the notation of \cite{pacejka_tire_book} we have:
\begin{subequations}
\begin{align} 
    F_x = {}& F_{x0} G_{xa}\label{eq:tire_forces_start}\\
    \begin{split}
        G_{xa} = {}& \cos\left( C_{xa} \arctan \left( B_{xa} \alpha \right.\right. \\
                   &\left.\left. - E_{xa}\left( B_{xa} \alpha - \arctan (B_{xa} \alpha) \right) \right) \right)
    \end{split}\\
    B_{xa} = {}& r_{Bx1} \cos(\arctan(r_{Bx2} \sigma))\\
    \begin{split}
        F_{x0} = {}& \mu N \sin\left(C_x \arctan\left(B_x \sigma \right.\right. \\
                   & \left.\left.- E_x (B_x \sigma - \arctan(B_x \sigma)) \right) \right)
    \end{split}\\
    F_y = {}& F_{y0} G_{ys}\\
    \begin{split}
        G_{ys} = {}& \cos\left( C_{ys} \arctan \left( B_{ys} \sigma \right.\right. \\
                   &\left.\left. - E_{ys}\left( B_{ys} \sigma - \arctan (B_{ys} \sigma) \right) \right) \right)
    \end{split}\\
    B_{ys} = {}& r_{By1} \cos(\arctan(r_{By2} \alpha))\\
    \begin{split}
        F_{y0} = {}& \mu N \sin\left(C_y \arctan \left(B_y \alpha  \right.\right. \\
                   & - \left.\left. E_y (B_y \alpha - \arctan(B_y\alpha))\right) \right)
    \end{split}\label{eq:tire_forces_end}
\end{align}
\end{subequations}
where $\sigma$ and $\alpha$ are the slip ratio and slip angle of the tire and $N$ the normal force on the tire. Parameters are provided in Table \ref{tab:vehicle_parameter_table} at the end of this paper. $F_x$ and $F_y$ are the lateral and longitudinal forces on the tire. We express this in the body frame in the form:
\begin{subequations}
\begin{equation}
    F^{t,b}_{ij,1} = N_{ij} \mu_{ij,1}(\sigma_{ij}, \alpha_{ij}, \gamma_{ij}) 
\end{equation}
\begin{equation}
    F^{t,b}_{ij,2} = N_{ij} \mu_{ij,2}(\sigma_{ij}, \alpha_{ij}, \gamma_{ij}) 
\end{equation}
\end{subequations}
where $_{ij}$ refers to any given tire and $\mu_{ij,1}$ and $\mu_{ij,2}$ capture remaining tire model terms, which are functions of $\mathcal{Z}$ and $\mathcal{U}$. 

We follow the slip definitions in \cite[pg. 67]{pacejka_tire_book}, whereby the slip angle $\alpha$ is 
\begin{equation}
    \alpha = \arctan \left(-\frac{V_{cy}}{V_{cx}^*}\right)
\end{equation}
here $V_{cy}$ is the lateral velocity of the tire at the surface of the road and $V_{cx}^*$ is the longitudinal velocity of the tire at the tire's effective radius. Both follow from translating the linear and angular of the vehicle to the rolling radius $r$ or effective radius $r_e$ of the tire and applying a rotation based on the steering angle of the tire. The expressions for this are straightforward and omitted here. Note that the effect of a curved surface is captured by nonzero $\omega^b_1$ and $\omega^b_2$ in this translation, a result of Equation \eqref{eq:angular_vel_constraint}.

We can now fill in the expressions for the net force and torque on the vehicle:
\begin{subequations}
\begin{equation}
    F^b_1 = F^{d,b}_1 + F^{g,b}_1 + F^{t,b}_1 
\end{equation}
\begin{equation}
    F^b_2 = F^{d,b}_2 + F^{g,b}_2 + F^{t,b}_2
\end{equation}
\begin{equation}
    F^b_3 = F^{d,b}_3 + F^{g,b}_3 + F^{t,b}_3 
\end{equation}
\begin{equation}
    K^b_1 = K^{d,b}_1 + h F^{t,b}_2  - N_{fr} t_f + N_{fl} t_f - N_{rr} t_r + N_{rl} t_r
\end{equation}
\begin{equation}
    K^b_2 = K^{d,b}_2 - h F^{t,b}_1 - (N_{fr} + N_{fl}) l_f + (N_{rr} + N_{rl}) l_r 
\end{equation}
\begin{equation}
    K^b_3 = K^{d,b}_3 + K^{t,b}_3
\end{equation}
\end{subequations}
where $F^{t,b}_1 = F^{t,b}_{fr,1} + F^{t,b}_{fl,1} + F^{t,b}_{rr,1} + F^{t,b}_{rl,1}$ and likewise for $F^{t,b}_2$. $K^{t,b}_3$ is the yaw torque produced by all longitudinal and lateral tire forces.

We now have all the expressions necessary to define $f(\mathcal{Z}, \mathcal{U}, \mathcal{G})$ from Equation \eqref{eq:dae_general_form}. Section \ref{sec:surface_motion_eqns} filled in the first three terms of $\dot{\mathcal{Z}}$, we can compute longitudinal and lateral tire forces from equations \eqref{eq:tire_forces_start} through \eqref{eq:tire_forces_end}, compute the resultant force and torque at the vehicle center of mass, add drag and gravity forces and use those expressions in \eqref{eq:ne_v1}, \eqref{eq:ne_v2} and \eqref{eq:ne_w3} along with constraints from Section \ref{sec:surface_motion_eqns}. This completes the first half of our DAE model. We obtain $g_c(\mathcal{Z}, \mathcal{U}, \mathcal{G})$ next by enforcing weight distribution constraints on $\mathcal{G}$.

\subsection{Weight Distribution}
We extend the approach of \cite{rucco2014development} to a vehicle on a nonplanar surface and derive the resulting force balance expressions. 

This involves four constraints to determine the four normal forces of each tire:
\begin{enumerate}
    \item Balance of the net normal force on the vehicle to remain in contact with the surface
    \item Longitudinal torque on the vehicle body 
    \item Lateral torque on the vehicle body
    \item A compatibility expression such as developed in \cite{rucco2014development}
\end{enumerate}
We use the compatibility expression of \cite{rucco2014development}:
\begin{equation}
    N_{fr} t_r - N_{fl} t_r - N_{rr} t_f + N_{rl} t_f = 0.
\end{equation}
As in \cite{rucco2014development} we make this constraint implicit by introducing the variables $N_f$, $N_r$ and $\Delta$ for which we have
\begin{subequations}
\begin{equation}
    N_{fr} = \frac{1}{2} N_f - t_f \Delta 
\end{equation}
\begin{equation}
    N_{fl} = \frac{1}{2} N_f + t_f \Delta 
\end{equation}
\begin{equation}
    N_{rr} = \frac{1}{2} N_r - t_r \Delta 
\end{equation}
\begin{equation}
    N_{rl} = \frac{1}{2} N_r + t_r \Delta.
\end{equation}
\end{subequations}

As a result, we choose
\begin{equation}
    \mathcal{G} = \left[ N_f, N_r, \Delta \right]
\end{equation}
for our DAE model \eqref{eq:dae_general_form}. It remains to find $g_c$. 

Since our DAE is in semi-explicit form we can find $g_c$ with the following approach:
\begin{enumerate}
    \item Compute $\dot{\mathcal{Z}}$ using $f$ and a guess for $\mathcal{G}$
    \item Find constraint forces and torques needed by $\dot{\mathcal{Z}}$
    \item Use these forces and torques to find a required $\hat{\mathcal{G}}$
    \item $g_c$ is then $\hat{\mathcal{G}}(\mathcal{Z}, \mathcal{U}, \mathcal{G}) - \mathcal{G} = 0$
\end{enumerate}

First we compute the required net force on the vehicle using the Newton Euler equations:
\begin{subequations}
\begin{equation}
    F^b_{3,\text{required}} = m \left( \dot{v}^b_3 + v^b_2 \omega^b_1 - v^b_1 \omega^b_2\right)
\end{equation}
\begin{equation}
    K^b_{1,\text{required}} = I^b_1 \dot{\omega^b_1} + (I^b_3 - I^b_2) \omega^b_2 \omega^b_3
\end{equation}
\begin{equation}
    K^b_{2,\text{required}} = I^b_2 \dot{\omega^b_2} + (I^b_1 - I^b_3) \omega^b_3 \omega^b_1
\end{equation}
\end{subequations}
where terms not part of $\mathcal{Z}$ can be computed from it (such as $\omega^b_1$) using expressions from Section \ref{sec:surface_motion_eqns}.

Next we remove drag forces, gravity forces and longitudinal/lateral tire forces to leave only the contributions of the normal force:
\begin{subequations}
\begin{equation}
    F^b_{3N,\text{required}} = F^b_{3,\text{required}} - F^{d,b}_3 - F^{g,b}_3
\end{equation}
\begin{equation}
    K^b_{1N,\text{required}} = K^b_{1,\text{required}} - K^{d,b}_1 - h F^{t,b}_2
\end{equation}
\begin{equation}
    K^b_{2N,\text{required}} = K^b_{2,\text{required}} - K^{d,b}_2 + h F^{t,b}_1.
\end{equation}
\end{subequations}

Finally, we impose these constraints to find required $N_f$, $N_r$ and $\Delta$:
\begin{subequations}
\begin{equation}
    N_{f,\text{required}} = \frac{l_r}{L}F^b_{3N,\text{required}} - \frac{1}{L} K^b_{2N,\text{required}}
\end{equation}
\begin{equation}
    N_{r,\text{required}} = \frac{l_f}{L}F^b_{3N,\text{required}} + \frac{1}{L} K^b_{2N,\text{required}}
\end{equation}
\begin{equation}
    \Delta_{\text{required}} = \frac{1}{2t_f^2 + 2t_r^2} K^b_{1N,\text{required}}.
\end{equation}
\end{subequations}
This gives us $g_c$, completing our two track vehicle model.

\subsection{Comparison Vehicle Models}
We compare the previous model to three others: 

First, the kinematic bicycle model of \cite{fork2021models} with an added friction cone constraint. We constrain the total acceleration of the vehicle from the tires to be less than $\frac{N\mu}{m}$, where $N$ is the total normal force on the vehicle, found in the same manner as in \cite{fork2021models}. Tire forces must balance three forces: the longitudinal acceleration input $a$, cornering force and the lateral component of gravity, which the tires must oppose to avoid the vehicle slipping sideways. These are combined to form the friction cone constraint. 

Second, we compare to a planar kinematic bicycle model - one in which the road surface has been projected to a flat plane, identical to the Frenet frame. 

Finally, we use a nonplanar dynamic bicycle model where we treat the longitudinal acceleration as an input but include lateral tire forces. We do so with quasi-static weight distribution: the total normal force is distributed between the front and rear wheels as it would be for a stationary vehicle on a flat surface. We also limit the acceleration input within $\pm \frac{N\mu}{m}$.

\section{SURFACE PARAMETERIZATION}\label{sec:surface_parameterization}
Thus far, no specific surface parameterization has been used, only knowledge of $\boldsymbol{x}^p_s$ and other partial derivatives of the surface. This allows our approach to capture different surfaces and parameterizations seamlessly. We now introduce a specific surface parameterization for use in numerical experiments.

We use and extend the Tait Bryan angle surface introduced in \cite{fork2021models}. This is a generalization of the Frenet Frame to arbitrary curvature, wherein the tangent $\boldsymbol{e}_s^c$, normal $\boldsymbol{e}_n^c$ and binormal $\boldsymbol{e}_y^c$ vectors of a curve in $\mathbb{R}^3$ are given by three Tait-Bryan angles. A parametric surface is then obtained by integrating the tangent vector to obtain a curve, and applying a linear binormal offset:
\begin{equation} \label{eq:general_centerline_ode}
    \frac{d \boldsymbol{x}^c}{d s} = \boldsymbol{e}_s^c(s)
\end{equation}
\begin{equation} \label{eq:general_centerline}
    \boldsymbol{x}^p(s,y) = \boldsymbol{x}^c(s) + y \boldsymbol{e}_y^c(s).
\end{equation}
More details on this may be found in \cite{fork2021models}. Note that $\boldsymbol{e}^c_n \neq \boldsymbol{e}^p_n$, the latter must be computed from $\boldsymbol{x}^p_s$ and $\boldsymbol{x}^p_y$.

We extend this surface parameterization by adding a normal offset, ie. the cross section of the road is no longer flat. Instead of \eqref{eq:general_centerline} we use the expression:
\begin{equation}
    \boldsymbol{x}^p(s,y) = \boldsymbol{x}^c(s) + y \boldsymbol{e}_y^c(s) + \boldsymbol{e}^c_n \frac{1}{\kappa}\left(1- \sqrt{1-y^2 \kappa^2} \right)
\end{equation}
which for $ \kappa \neq 0$ and $|y| \leq \frac{1}{\kappa} $ gives us an arc segment. This is illustrated later in Figure \ref{fig:tube_turn}.

With this normal offset, the surface parameterization is no longer orthogonal, complicating computation of $\mathbf{J}$. In \cite{fork2021models} we introduced a method to compute $\mathbf{J}$ for orthogonal surface parameterizations using $\theta^s$. Here we introduce the general method:

\begin{subequations}
\begin{align}
    \theta^p =& - \sin^{-1} \left( \frac{\boldsymbol{x}^p_s \cdot \boldsymbol{x}^p_y}{||\boldsymbol{x}^p_s||~||\boldsymbol{x}^p_y||} \right)\\
    \mathbf{J} =& 
    \begin{bmatrix}
        \cos(\theta^s) ||\boldsymbol{x}^p_s|| & -\sin(\theta^s) ||\boldsymbol{x}^p_s||\\
        \sin(\theta^s - \theta^p)||\boldsymbol{x}^p_y|| & \cos(\theta^s - \theta^p)||\boldsymbol{x}^p_y||
    \end{bmatrix}
\end{align}
\end{subequations}
where the only difference from \cite{fork2021models} is the added $\theta^p$, which is an angular measure of how far the surface parameterization is from orthogonal.

\section{OPTIMAL CONTROL} \label{sec:optimal_control}
\subsection{Vehicle Models}
We use our two track and comparison models to compute minimum time trajectories on nonplanar surfaces. For bicycle models we limit the net normal force on the vehicle: 
\begin{equation}
    0 \leq N \leq N_{max}
\end{equation}
and limit individual tire forces for the two track model:
\begin{equation}
    0 \leq N_{ij} \leq \frac{1}{2} N_{max}
\end{equation}
where we use $\frac{1}{2}$ instead of $\frac{1}{4}$ to reflect a need for overconservatism in the simpler models\footnote[1]{With the simpler models, all force could be distributed on the front wheels without the model capturing this}. 

\subsection{Optimization Framework}
We used collocation~\cite[ch. 10]{biegler_book} to set up a raceline optimization problem using our model. Collocation is widely used for raceline computation \cite{christ2021time, 3d_part_1}; we refer the reader to the previous three references for more information. We used $7^{th}$ order Gauss-Legendre collocation with $100$ intervals of uniform path length along the length of the track. To improve numerical performance we normalized the algebraic state $\mathcal{G}$ by $mg$. We penalized the lap time of the raceline with quadratic regularization applied to input and input rate: $J = \int(1 + u^TRu + \dot{u}^TR_d \dot{u})dt$. We used $R = R_d = 0.001I$, where $I$ is an appropriately sized identity matrix. For closed tracks we enforced a loop closure constraint on state and input variables, otherwise we fixed the initial state of the vehicle.

\section{RESULTS} \label{sec:results}
We implemented all vehicle models, the previously described surface parameterization and the raceline problem symbolically in CasADi~\cite{Andersson2018}, which was then solved using IPOPT~\cite{ipopt} with the linear solver MUMPS~\cite{MUMPS_1}. All programs were run and timed on an AMD Ryzen 5700U CPU at 4.3GHz. 

\begin{figure}
    \centering
    \includegraphics[width=0.95\linewidth]{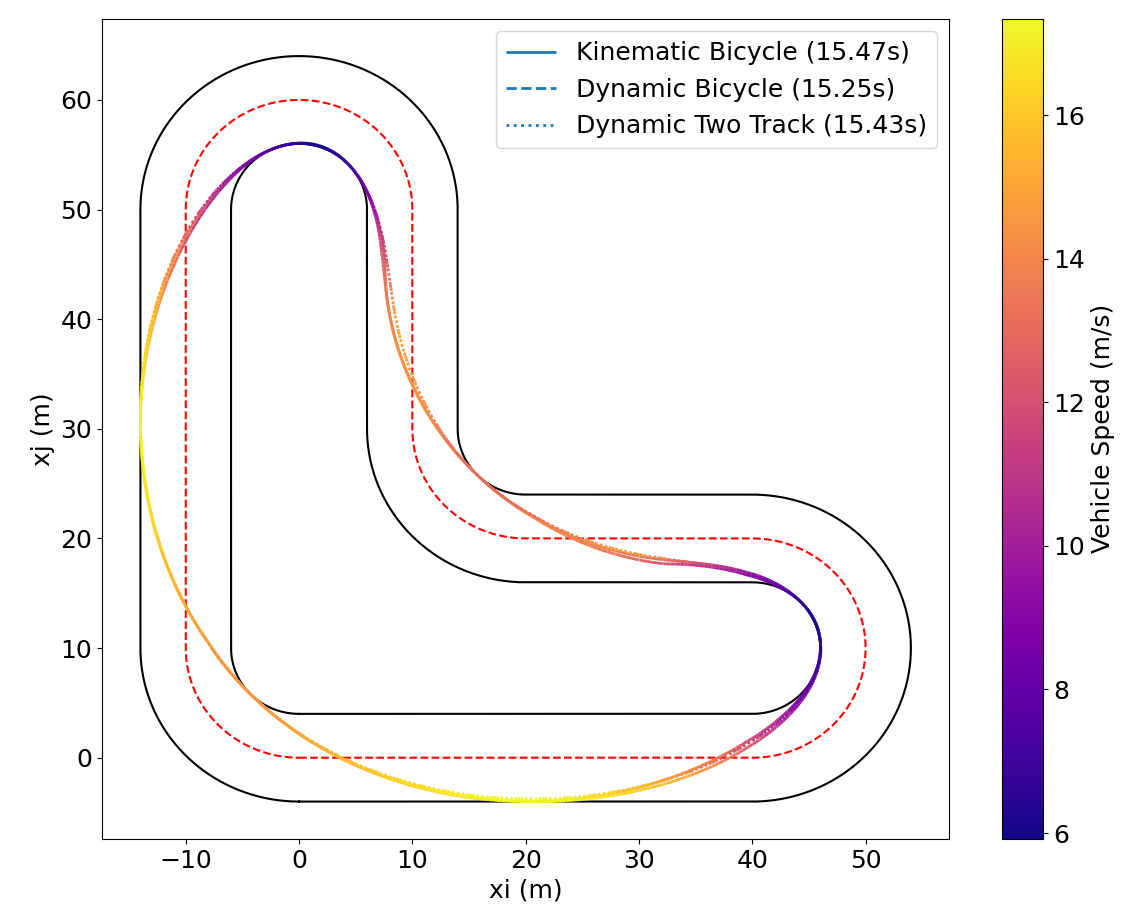}
    \caption{Planar validation racelines}
    \label{fig:l_track}
\end{figure}

To validate our models relative to one another we compared them on a flat, L-shaped track (Figure \ref{fig:l_track}). All three models produced visually identical racelines and lap times as seen in Figure \ref{fig:l_track}, solve times and the time required to set up the NLP in CasADi are reported in Table \ref{tab:raceline_solve_times_L}. We omit the planar kinematic bicycle model as it is identical to the general kinematic bicycle model for this flat surface. 

\begin{figure*}
    \centering
    \includegraphics[width=0.85\linewidth]{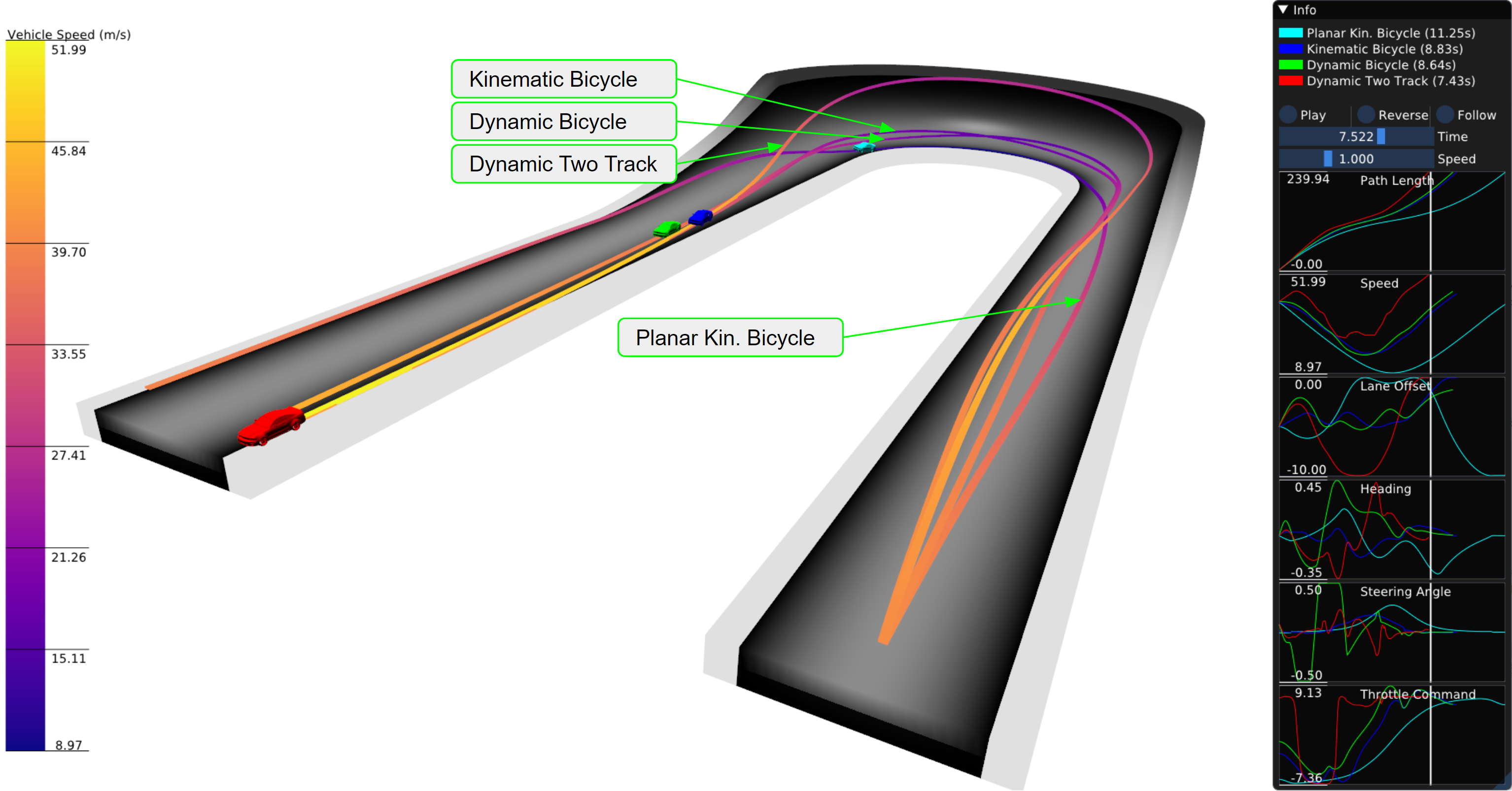}
    \caption{Nonplanar racelines for each vehicle model. Racelines are colored by speed whereas vehicles are colored by model, with the duration of each raceline in the legend.}   
    \label{fig:tube_turn}
\end{figure*}

Next we computed racelines on a tube-shaped turn (Figure \ref{fig:tube_turn}). Here the two track vehicle model achieved a faster lap time than the other two models by leveraging the curved turn to maintain vehicle speed throughout. Racelines are shown in Figure \ref{fig:tube_turn} with the color of each car corresponding to the model used. Solve times and the time required to set up the NLP in CasADi for the nonplanar models are reported in Table \ref{tab:raceline_solve_times_tube}. An animated video of the nonplanar racelines can be found at \href{https://youtu.be/4nCYGlKpd2A}{https://youtu.be/4nCYGlKpd2A}

\begin{table}
\caption{Raceline Solve Times: Planar Track} \label{tab:raceline_solve_times_L}
\begin{center}
\begin{tabular}{|c|c|c|c|}
\hline
\textbf{Time} (s) & \textbf{Kinematic} & \textbf{Dynamic} & \textbf{Two-Track} \\
\hline 
Setup     & 3.23 & 6.14 & 21.0 \\
IPOPT     & 0.49 & 7.07 & 11.0 \\
NLP Evals & 0.14 & 1.35 & 7.98 \\
\hline
\end{tabular}
\end{center}
\vspace{-4mm}
\end{table}

\begin{table}
\caption{Raceline Solve Times: Nonplanar Track} \label{tab:raceline_solve_times_tube}
\begin{center}
\begin{tabular}{|c|c|c|c|c|}
\hline
\textbf{Time} (s) & \multicolumn{2}{c|}{\textbf{Kinematic}} & \textbf{Dynamic} & \textbf{Two-Track} \\
\hline Setup & 3.72 & 10.5 & 13.3 & 34.9 \\
IPOPT        & 1.33 & 2.21 & 14.3 & 37.8 \\
NLP Evals    & 0.29 & 1.57 & 7.41 & 28.7 \\
\hline
\end{tabular}
The first Kinematic column is the planar model.
\end{center}
\vspace{-4mm}
\end{table}

\section{CONCLUSION} \label{sec:conclusion}
In this paper we developed a two track vehicle model suitable for optimal control on nonplanar surfaces. In contrast to prior work it can be applied systematically to arbitrary nonplanar surfaces and is computationally tractable for model-based control. We demonstrated the use of our model for computing time optimal trajectories on a nonplanar surface, where it outperformed simpler nonplanar vehicle models and a planar vehicle model. 

\renewcommand{\refname}{REFERENCES}
\printbibliography

\begin{appendix}
\section{Parameters}
\begin{table}[htbp] 
\caption{Vehicle Parameters (MKS where applicable)} \label{tab:vehicle_parameter_table}
\begin{center}
\begin{tabular}{|c|c|c|c|}
\hline
\textbf{Parameter} & \textbf{Value} & \textbf{Parameter} & \textbf{Value} \\
\hline 
$B_x$ & 16 & $B_y$ & 13\\
$C_x$ & 1.58 & $C_y$ & 1.45\\
$E_x$ & 0.1 & $E_y$ & -0.8\\
\hline 
$E_{xa}$ & -0.5 & $E_{ys}$ & 0.3\\
$C_{xa}$ & 1 & $C_{ys}$ & 1\\
$r_{Bx1}$ & 13 & $r_{By1}$ & 10.62\\
$r_{Bx2}$ & 9.7 & $r_{By2}$ & 7.82\\

\hline 
$\mu $ & 0.75 & & \\
$r$ & 0.3  & $r_e$ & 0.3 \\
\hline
$m$  & 2303  & $g$  & 9.81 \\
$l_f$ & 1.52  & $l_r$ & 1.50  \\
$t_f$ & 0.625  & $t_r$ & 0.625  \\
$h$  & 0.592  & $I_1^b$ & 956  \\
$I_2^b$ & 5000  & $I_3^b$ & 5520 \\
\hline
$\gamma$ & $\in [-.5,.5] \si{\radian}$ & $N_{max}$ & 40 \si{\kilo\newton} \\
\hline
\end{tabular}
\label{tab1}
\end{center}
\vspace{-4mm}
\end{table}

\section{Kinematic Bicycle Friction Cone}
As we limit ourselves to tire forces that are linear in normal force, we use a friction cone constraint to enforce that the acceleration magnitude of the bicycle model is less than
\begin{equation}
    \frac{\mu N}{m}
\end{equation}
where once more $N$ is the net normal force on the vehicle. 

Three contributions of acceleration are present:
\begin{enumerate}
    \item The longitudinal acceleration input $a$
    \item Lateral acceleration necessary for cornering
    \item Lateral acceleration necessary to avoid sliding on a banked surface
\end{enumerate}
Note that that the effect of a sloped surface is captured by altering the expression for $\dot{v}$, the rate of change of the speed of the vehicle, rather than being part of the tire friction cone constraint.

We model lateral acceleration with the expression
\begin{equation}
    \frac{v^2 \gamma}{L}
\end{equation}
where $\gamma$ is the front wheel steering angle, $L$ the wheelbase length and their ratio approximates the curvature of the trajectory of the vehicle. 

Meanwhile the lateral acceleration necessary on a banked surface is the component of gravity orthogonal to the direction the vehicle is moving. The components of gravity can be found from 3D vehicle orientation, which we derive expressions for in Appendix \ref{app:orientation}. By definition, a kinematic bicycle moves in direction
$ \boldsymbol{e}^b_1 \cos \beta + \boldsymbol{e}^b_2 \sin \beta $. With our choices of sign convention, the cornering acceleration to oppose gravity is:
\begin{equation}
    a^g_t = -mg \left( \boldsymbol{e}^b_1 \cdot \boldsymbol{e}^g_3 \sin \beta -  \boldsymbol{e}^b_2 \cdot \boldsymbol{e}^g_3 \cos \beta \right)
\end{equation}
our friction cone constraint is then
\begin{equation}
    \left( \frac{m}{\mu N}\right)^2 \left(a^2 + \left(a^g_t + \frac{v^2 \gamma}{L}\right)^2 \right) \leq 1
\end{equation}
where we have replaced the Euclidean magnitude with its square for numerical performance benefits, and divided by the squared allowable acceleration to achieve a constraint with constant bound. 

\section{Computing Vehicle Orientation} \label{app:orientation}
Though we describe vehicle pose with the single variable $\theta^s$, it is possible to compute the full 3D pose of the vehicle and to do so from only $\theta^s$ and the properties of the parametric surface. Besides being useful for visualization purposes, this is essential to compute gravity components for vehicle models. Furthermore, without describing orientation using only $\theta^s$, we would need full 3D orientation, a rotation matrix, to be a decision variable in model-based control. Here we introduce the general method to compute 3D orientation from $\theta^s$ by computing $\boldsymbol{e}^b_1$, $\boldsymbol{e}^b_2$ and $\boldsymbol{e}^b_3$ from $\theta^s$ and first derivatives of the parametric surface. 

First, and trivially, we have $\boldsymbol{e}^b_3 = \boldsymbol{x}^p_n$, otherwise the vehicle would not be tangent to the surface. 

Second we consider the expression:
\begin{subequations}
\begin{equation}
    \boldsymbol{e}^b_1 = \alpha \boldsymbol{x}^p_s + \beta \boldsymbol{x}^p_y
\end{equation}
\begin{equation}
    \boldsymbol{e}^b_2 = \gamma \boldsymbol{x}^p_s + \delta \boldsymbol{x}^p_y
\end{equation}
\end{subequations}
where $\alpha$, $\beta$, $\gamma$ and $\delta$ are unknowns to solve for. We first split each equation into two equations by taking inner products with respect to $\boldsymbol{x}^p_s$ and $\boldsymbol{x}^p_y$:
\begin{subequations}
\begin{equation}
    \begin{bmatrix}
        \boldsymbol{e}^b_1 \cdot \boldsymbol{x}^p_s \\ 
        \boldsymbol{e}^b_1 \cdot \boldsymbol{x}^p_y 
    \end{bmatrix}
    = 
    \mathbf{I}
    \begin{bmatrix}
        \alpha \\ \beta
    \end{bmatrix}
\end{equation}
\begin{equation}
    \begin{bmatrix}
        \boldsymbol{e}^b_2 \cdot \boldsymbol{x}^p_s \\ 
        \boldsymbol{e}^b_2 \cdot \boldsymbol{x}^p_y 
    \end{bmatrix}
    = 
    \mathbf{I}
    \begin{bmatrix}
        \gamma \\ \delta
    \end{bmatrix}.
\end{equation}
\end{subequations}
We can merge and solve the two equations by recognizing that they are the columns of $\boldsymbol{J}$:
\begin{equation}
    \begin{bmatrix}
        \alpha & \gamma \\ \beta & \delta
    \end{bmatrix}
    = \mathbf{I}^{-1} \mathbf{J}
\end{equation}
Therefore we have (vector symbols treated as column vectors):
\begin{equation}
    \begin{bmatrix}
        \boldsymbol{e}^b_1 & \boldsymbol{e}^b_2
    \end{bmatrix}
    = \begin{bmatrix}
        \boldsymbol{x}^p_s & \boldsymbol{x}^p_y
    \end{bmatrix}
    \mathbf{I}^{-1} \mathbf{J}
\end{equation}
\end{appendix}
which we can compute knowing only $\theta^s$ and first order knowledge of the parametric surface.

\end{document}